\documentclass[fleqn,usenatbib]{mnras}
\usepackage[T1]{fontenc}
\usepackage{ae,aecompl}
\usepackage{graphicx}
\usepackage{amsmath}
\usepackage{amssymb}
\usepackage{txfonts}
\usepackage{cleveref}
\usepackage{bm}
\usepackage{hyperref}
\usepackage{tensor}
\usepackage{microtype}
\usepackage{mathtools}

\usepackage{cleveref}
\usepackage{upgreek}
\usepackage{xcolor}

\graphicspath{./}

\newcommand{\ltsima}{$\; \buildrel < \over \sim \;$}
\newcommand{\lsim}{\lower.5ex\hbox{\ltsima}}
\newcommand{\gtsima}{$\; \buildrel > \over \sim \;$}
\newcommand{\gsim}{\lower.5ex\hbox{\gtsima}}

\newcommand{\revision}[1]{\textcolor{black}{#1}}

\title[Equivalence principle wich located FRBs]
{Consistent Constraints on the Equivalence Principle from localised Fast Radio Bursts}
\author[Reischke, Hagstotz]
{
Robert Reischke\thanks{E-mail:  \href{mailto:reischke@astro.ruhr-uni-bochum.de}{reischke@astro.ruhr-uni-bochum.de}}$^{1}$ and Steffen Hagstotz\thanks{E-mail: \href{mailto:steffen.hagstotz@lmu.de}{steffen.hagstotz@lmu.de}}$^{2,3}$
\\
$^1$ Ruhr University Bochum, Faculty of Physics and Astronomy, Astronomical Institute (AIRUB),\\ \hspace{0.15cm} German Centre for Cosmological Lensing, 44780 Bochum, Germany
\\
$^2$  Universitäts-Sternwarte, Fakultät für Physik, Ludwig-Maximilians Universität München, \\
Scheinerstraße 1, D-81679 München, Germany
 \\
$^3$  Excellence Cluster ORIGINS, Boltzmannstraße 2, D-85748 Garching, Germany
}

\begin{document}
\pagerange{\pageref{firstpage}--\pageref{lastpage}}
\pubyear{2022}
\maketitle
\label{firstpage}

\begin{abstract}
Fast Radio Bursts (FRBs) are short astrophysical transients of extragalactic origin. Their burst signal is dispersed by the free electrons in the large-scale-structure (LSS), leading to delayed arrival times at different frequencies. Another potential source of time delay is the well known Shapiro delay, which measures the space-space and time-time metric perturbations along the line-of-sight. If photons of different frequencies follow different trajectories, i.e. if the universality of free fall guaranteed by the weak equivalence principle (WEP) is violated, they would experience an additional relative delay. This quantity, however, is not an observable on the background level as it is not gauge independent, which has led to confusion in previous papers. Instead, an imprint can be seen in the correlation between the time delays of different pulses.
In this paper, we derive robust and consistent constraints from twelve localised FRBs on the violation of the WEP in the energy range between 4.6 and 6 meV. In contrast to a number of previous studies, we consider our signal to be not in the model, but in the covariance matrix of the likelihood. To do so, we calculate the covariance of the time delays induced by the free electrons in the LSS, the WEP breaking terms, the Milky Way and host galaxy. By marginalising over both host galaxy contribution and the contribution from the free electrons, we find that the parametrised post-Newtonian parameter $\gamma$ characterising the WEP violation must be constant in this energy range to 1 in $10^{13}$ at 68$\;\%$ confidence. These are the tightest constraints to-date on $\Delta\gamma$ in this low energy range. 
\end{abstract}

\begin{keywords}
cosmology: theory, large-scale structure of Universe, radio continuum:  transients
\end{keywords}

\section{Introduction}
\renewcommand{\arraystretch}{1.25}

\begin{table*}
    \centering
    \begin{tabular}{cccccc}
       Name  & DM $[\mathrm{pc}\;\mathrm{cm}^{-3}]$ & $\mathrm{DM}_\mathrm{MW}$ $[\mathrm{pc}\;\mathrm{cm}^{-3}]$ &$\nu$ $[\mathrm{MHz}]$  & $z$ & $\mathrm{d}\nu$ $[\mathrm{MHz}]$
       \\
       \hline\hline    
         20191001 & 506.92 & 44.2  & 919.5  & 0.23  & 336\\
         20200430 & 380.1  & 27.0    & 864.5 & 0.161   & 336  \\
20200906 & 577.8  & 35.9  & 864.5 & 0.36879 & 336\\
20180924$^*$   & 362.4  & 40.5 & 1297.5 &  0.3214 & 336 \\
20181112$^*$   & 589.0  & 40.2 & 1297.5 &  0.4755 & 336 \\
20190102$^*$   & 364.5  & 57.3 & 1271.5 & 0.291 & 336 \\
20190608$^*$   & 339.5  & 37.2 & 1271.5 & 0.1178 & 336 \\
20190611.2$^*$ & 322.2  & 57.6 & 1271.5  & 0.378 & 336 \\
20190711$^*$   & 594.6  & 56.6 & 1271.5 & 0.522 & 336 \\
20190714$^*$   & 504.7  & 38.5 & 1271.5 & 0.209 & 336 \\
20191228$^*$   & 297.5  & 32.9 & 1271.5 & 0.243 & 336 \\
20190523  & 760.8 & 47 & 1411 & 0.66 & 225 \\
    \end{tabular}
    \caption{FRBs used in this work. Given is the FRB name, the observed DM, the estimated DM from the MW (also often referred to as the ISM component), the central frequency, the redhsift from the host identification and the bandwidth. The FRBs are taken from \citet{bhandari_limits_2020, bhandari_characterizing_2022, heintz_host_2020,bannister_single_2019, prochaska_low_2019,macquart_census_2020} and \citet{ravi_fast_2019}. FRBs marked with $^*$ will be used in a sub-sample analysis as described in \Cref{sec:results}.}
    \label{tab:frb_set}
\end{table*}

\label{sec:intro}

Fast Radio Bursts (FRBs) are very short transients lasting only a few milliseconds and cover a frequency range from a few hundred to a few thousand MHz. The scattering with free electrons in the ionised intergalactic medium (IGM) disperses the pulse, leading to a typical time delay $\Delta t \propto \nu^{-2}$. Its amplitude is called the dispersion measure (DM) \citep[see e.g.][]{thornton_population_2013, petroff_real-time_2015, connor_non-cosmological_2016, champion_five_2016,chatterjee_direct_2017,macquart_census_2020}. FRBs have sparked a flurry of research over the past years and their potential is massive, as discussed in the review by \citet{petroff_fast_2022}. The DM is proportional to the integrated electron density along the line-of-sight between the source and the observer and forms a unique astrophysical and cosmological probe. Even though the mechanism of the bursts is still under debate,\footnote{For a compilation of some proposed mechanisms for FRBs, see \href{https://frbtheorycat.org}{https://frbtheorycat.org} \citep{platts_living_2019}.}, their isotropic distribution across the sky and large observed DM advocates an extragalactic origin \revision{\citep[allthough some might also be galactic, see][]{andersen_bright_2020}}. Thus, the DM can test the distribution of diffuse electrons in the large-scale structure (LSS). 

FRBs have recently been proposed to test the weak equivalence principle (WEP). The WEP guarantees the universality of free fall, one of the key axioms of General Relativity. When the WEP is broken, photons of different frequency, i.e. energy, can follow different null-geodesics. This is also true for other light (relativistic) particles  such as neutrinos \citep[see][]{bose_effect_1988}. In effect, pulses from short transients would pick up a massive time delay in their signal in time-frequency space if the WEP is broken due to the cosmological distances involved. 
Accordingly, transients at cosmological distances are a promising ground to test the WEP. 

There are two components to the DM, the homogeneous (background) and the inhomogeneous component sourced by perturbations. While the former is only accessible through FRBs with known host, since they provide an independent redshift estimate \citep{zhou_fast_2014,walters_future_2018,hagstotz_new_2022,macquart_census_2020,wu_8_2022,james_measurement_2022}, the latter can be studied with the full FRB sample through correlations \citep{masui_dispersion_2015,shirasaki_large-scale_2017,rafiei-ravandi_chimefrb_2021,bhattacharya_fast_2020,takahashi_statistical_2021}. 
In principle, any breaking of the WEP would immediately lead to a much higher DM \revision{correlations} than expected, thus opening a window to put tight constraints on the WEP. While there are a number of studies using FRBs with host identification to constrain the WEP, it was pointed out in \citet{minazzoli_shortcomings_2019} and \citet{reischke_consistent_2022} that these constraints are not accurate since they all assume a form of the Shapiro delay derived from a metric with weak perturbations that vanish at infinity. In cosmology, however, potentials do not vanish at spatial infinity (unless they vanish everywhere, rendering the whole discussion moot) due to the symmetries of the Friedmann-Robertson-Walker metric. \citet{reischke_consistent_2022} suggest a way out of this dilemma by using angular statistics of FRBs instead, which yields a well defined equation for the Shapiro time delay fluctuations \citep[for fluctuations of the DM see e.g.][]{masui_dispersion_2015,shirasaki_large-scale_2017,rafiei-ravandi_characterizing_2020,reischke_probing_2021,bhattacharya_fast_2020,takahashi_statistical_2021,rafiei-ravandi_chimefrb_2021}, which is, in contrast to the classical approach, gauge invariant. This was partially already applied to gamma ray bursts \citep{bartlett_constraints_2021} using full forward modelling by combining realisations of the local density field which found $\Delta\gamma < 2.1\times 10^{-15}$.

In this paper, we intend to revisit FRBs with host identification as promising tools to test the WEP. Recently, \citet{reischke_covariance_2023} have calculated the covariance matrix from the LSS for DM$-z$ observations. Furthermore, \citet{nusser_testing_2016} already used individual FRBs to constrain $\Delta\gamma$ from the fluctuations. We use the full covariance and apply it to a current data set of FRBs with host identification to put constraints on $\Delta\gamma$. In this way, the parameter dependence does not lie in the single dispersion itself, but in the covariance.

\section{Shapiro delay tests with localised Fast Radio Bursts}
\label{sec:shapiro_Frbs}
\subsection{Using FRBs to test the equivalence principle}
The observed time delay, $\Delta t_\mathrm{obs}$, between different frequency bands of an astrophysical transient can be split into several contributions:
\begin{equation}
    \Delta t_\mathrm{obs} = \Delta t_\mathrm{int} +  \Delta t_\mathrm{grav}\;.
\end{equation}
$ \Delta t_\mathrm{int}$ is the intrinsic time delay due to the source and the type of transient. In the case of FRBs this can be split into the DM contribution $ \Delta t_\mathrm{DM}$ and a potential source contribution $ \Delta t_\mathrm{s}$ which we assume to vanish. 
With this we are left with
\begin{equation}
\label{eq:observed_time_delay}
 \Delta t_\mathrm{obs} = \Delta t_s + \Delta t_\mathrm{DM} +  \Delta t_\mathrm{grav}\;,    
\end{equation}
where the last term is the difference in the gravitational time delay, between photons of different frequencies.
We write the weakly perturbed Friedman-Robertson-Walker (FRW) line element in conformal Newtonian gauge within the PPN formulation \citep{will_confrontation_2014} as follows:
\begin{equation}
    \label{eq:metric}
    \mathrm d s^2 = -\biggl(1+ \frac{2 \phi}{c^2} \biggr)  \,\mathrm c^2 \mathrm{d} t^2 + a^2(t) \; \biggl(1 - \frac{2 \gamma \phi}{c^2} \biggr) \, \mathrm d \mathbf x^2\;,
\end{equation}
with the gauge potential $\phi${, the scale factor $a$ and the comoving coordinates $\mathbf{x}$}. The PPN parameter $\gamma$ measures the deviation from the Newtonian expectation, with $\gamma=1$ in general relativity. The time delay experienced by a photon is then given by: 
\begin{equation} 
\label{eq:cosmo_shapiro_1}
   t_\mathrm{grav}(\hat{\boldsymbol{x}}) = \frac{1+\gamma}{c^3}\int^{\chi_\mathrm{s}}_{0} \mathrm{d}\chi\; a(\chi) \, \phi \big( \hat{\boldsymbol{x}}\chi,a(\chi) \big) \;,
\end{equation}
where $\chi$ is the comoving distance at the background level. Considering two photons at two different frequencies $\nu_{1,2}$ this turns into
\begin{equation} 
\label{eq:cosmo_shapiro}
   \Delta t_\mathrm{grav}(\hat{\boldsymbol{x}}) = \frac{\Delta\gamma_{1,2}}{c^3}\int^{\chi_\mathrm{s}}_{0} \mathrm{d}\chi\; a(\chi) \, \phi \big( \hat{\boldsymbol{x}}\chi,a(\chi) \big) \;,
\end{equation}
where $\Delta\gamma_{1,2}$ measures by how much the time delay changes between frequencies $\nu_{1,2}$. If the WEP holds, one expects $\Delta\gamma_{1,2} = 0$ due to the universality of free fall.
As discussed in \citet{reischke_consistent_2022} this expression is not plagued by any divergences and respects the cosmological symmetry assumptions by construction. Time delays picked up by photons along the line-of-sight can be both positive or negative. Hence, it is impossible to use individual FRBs to constrain the WEP unless the potential along the line of sight is known.

\subsection{Dispersion Measure and Time Delay Statistics}
\label{sec:dispersion_measure_statistics}
The observed time delay in the direction $\hat{\boldsymbol{x}}$ of a source at redshift $z$ is interpreted as an observed DM
\begin{equation}
\label{eq:delay_measure}
    \Delta t_\mathrm{obs}(\hat{\boldsymbol{x}},z) \propto \mathrm{DM}_\mathrm{obs}(\hat{\boldsymbol{x}}, z) \nu^{-2}\;.
\end{equation}
In particular, the time delay in a frequency band bounded by$\nu_{1,2}$ is
\begin{equation}
\begin{split}
    \Delta t_\mathrm{obs}(\hat{\boldsymbol{x}},z) = &\ t_{\nu_1,\mathrm{obs}}(\hat{\boldsymbol{x}},z) - t_{\nu_2,\mathrm{obs}}(\hat{\boldsymbol{x}},z) \\ = &\ \mathcal{K} \, \mathrm{DM}_\mathrm{obs}(\hat{\boldsymbol{x}},z) \, \left(\nu^{-2}_1-\nu^{-2}_2\right)\;,
\end{split}
    \end{equation}
where we absorb all the constants in $\mathcal{K} = e^2/(2\uppi m_\mathrm{e} c)$ and for consistency assume $\nu_2 >\nu_1$.
Here, $e$ and $m_\mathrm{e}$ denote the charge and mass of an electron, respectively. WEP breaking now leads to a shift in the time delay as such:
\begin{equation}
\label{eq:dispersion_measure_observed}
    \mathrm{DM}_\mathrm{obs}(\hat{\boldsymbol{x}},z) \to \mathrm{DM}_\mathrm{obs}(\hat{\boldsymbol{x}},z) + \mathcal{D}_\mathrm{grav}(\hat{\boldsymbol{x}},z)\;,
\end{equation}
where $\mathcal{D}_\mathrm{grav}(\hat{\boldsymbol{x}},z)$ is the time delay from \Cref{eq:cosmo_shapiro} interpreted as a DM in direction $\hat{\boldsymbol{x}}$ and up to redshift $z$:
 \begin{equation}
 \label{eq:dispersion_measure_grav}
    \mathcal{D}_\mathrm{grav}(\hat{\boldsymbol{x}},z) = \frac{\Delta\gamma_{1,2}}{\mathcal{K}c^3\left(\nu^{-2}_1-\nu^{-2}_2\right)}\int_0^{\chi(z)}\mathrm{d}\chi^\prime\; a(\chi^\prime) \, \phi\big(\hat{\boldsymbol{x}}\chi^\prime,z(\chi^\prime)\big)\;.
\end{equation}
As mentioned before, this contribution can be positive and negative, a problem which has not been addressed in previous studies. Note that this identification is subject to the $\nu^{-2}$ law, thus providing a preferred frequency shape of the \revision{WEP}-breaking term. However, the null hypothesis is $\Delta\gamma = 0$, as predicted by GR, and any additional contribution will immediately show up in the inferred DM budget. 

The non-gravitational contribution in \Cref{eq:dispersion_measure_observed} is  split into three parts:
\begin{equation}
\label{eq:dispersion_measure_contributions}
    \mathrm{DM}_\mathrm{obs}(\hat{\boldsymbol{x}},z) = \mathrm{DM}_\mathrm{LSS}(\hat{\boldsymbol{x}},z) + \mathrm{DM}_\mathrm{MW}(\hat{\boldsymbol{x}}) + \mathrm{DM}_\mathrm{host}(z)\;.
\end{equation}
The contribution from the Milky Way, $\mathrm{DM}_\mathrm{MW}(\hat{\boldsymbol{x}})$, models of the galactic electron distribution predict $\mathrm{DM}_\mathrm{MW} \lsim 50 \; \mathrm{pc}\;\mathrm{cm}^{-3}$ \citep{yao_new_2017}, unless the burst is in the galactic plane. Here we assume that this contribution can be modelled and subtracted from the signal. It will, however, induce additional scatter in the observed signal. 
For the host galaxy contribution $\mathrm{DM}_\mathrm{host}$, the situation is less clear and we will discuss this in more detail in \Cref{sec:results}. Finally, the LSS contribution is the line-of-sight integral over the electron distribution:
\begin{equation}
\label{eq:dispersion_measure_general}
    \mathrm{DM}_\mathrm{LSS}(\hat{\boldsymbol{x}},z) = \int_0^z \! n_\mathrm{e}^\mathrm{\revision{IGM}}({\boldsymbol{x}},z^\prime) \, \frac{1+z^\prime}{H(z^\prime)} \, \mathrm{d}z^\prime\;.
\end{equation}
Here $H(z)= H_0E(z)$ is the Hubble function, and {$n_\mathrm{e}^\mathrm{IGM}$ the number density of electrons in the IGM}, which can be related to the electron density contrast $\delta_\mathrm{e}(\boldsymbol{x},z)$:
\begin{equation}
\label{eq:electron_number_density}
    n_\mathrm{e}^\mathrm{\revision{IGM}}({\boldsymbol{x}},z)
    = \revision{F_\mathrm{IGM}(z) \,} \frac{\bar\rho_\mathrm{b}(z)}{m_\mathrm{p}}\left[1+ \delta_\mathrm{e}({\boldsymbol{x}},z)\right] \; ,
\end{equation}
with the {mean baryon mass density $\bar\rho_\mathrm{b}(z)$},
the proton mass $m_\mathrm{p}$ and the fraction of electrons in the IGM, $F_\mathrm{{IGM}}(z)$, which can be expressed as follows:
\begin{equation}
    \label{eq:ionization_fraction_of_the_igm}
    F_\mathrm{\revision{IGM}}(z) = f_\mathrm{IGM}(z) \, \biggl[Y_\mathrm{H}X_{\mathrm{e},\mathrm{H}}(z) + \revision{\frac{1}{2}} Y_\mathrm{He}X_{\mathrm{e},\mathrm{He}}(z)\biggr]\;.
\end{equation}
Here $Y_\mathrm{H} = 0.75$ and $Y_\mathrm{He} = 0.25$ are the mass fractions of hydrogen and helium, respectively, $X_{\mathrm{e},\mathrm{H}}(z)$ and $X_{\mathrm{e},\mathrm{He}}(z)$ are their ionization fractions, and $f_\mathrm{IGM}(z)$ is the mass fraction of baryons in the IGM. We assume $X_{\mathrm{e},\mathrm{H}} =X_{\mathrm{e},\mathrm{He}} = 1$ and $f_\mathrm{IGM}(z) = 90\%\; (80\%)$ at $z \gsim 1.5 \; (\lsim 0.4)$ 
\citep{meiksin_physics_2009,becker_detection_2011,shull_baryon_2012} with a linear interpolation. By rearranging \Cref{eq:electron_number_density} {in terms of today's dimensionless baryon density parameter $\Omega_{\mathrm{b}0}$ one finds:
\begin{equation}
\label{eq:dispersion_measure_specific}
    \mathrm{DM}_\mathrm{LSS}(\hat{\boldsymbol{x}},z) = \frac{3H_0^2\Omega_{\mathrm{b}0}\chi_H}{8\uppi G m_\mathrm{p}}\int_0^z \mathrm{d}z^\prime \frac{1+z^\prime}{E(z^\prime)}F_\mathrm{\revision{IGM}}(z^\prime)[1+\delta_\mathrm{e}({\boldsymbol{x}},z^\prime)]\;,
\end{equation}
and we absorb the total amplitude into a prefactor
\begin{equation}
\label{eq:prefactor}
    \mathcal{A} = \frac{3H_0^2\Omega_{\mathrm{b}0}\chi_H}{8\uppi G m_\mathrm{p}}\;,
\end{equation}
where $\chi_H$ is the Hubble radius today.

\begin{figure}
    \centering
    \includegraphics[width = .47\textwidth]{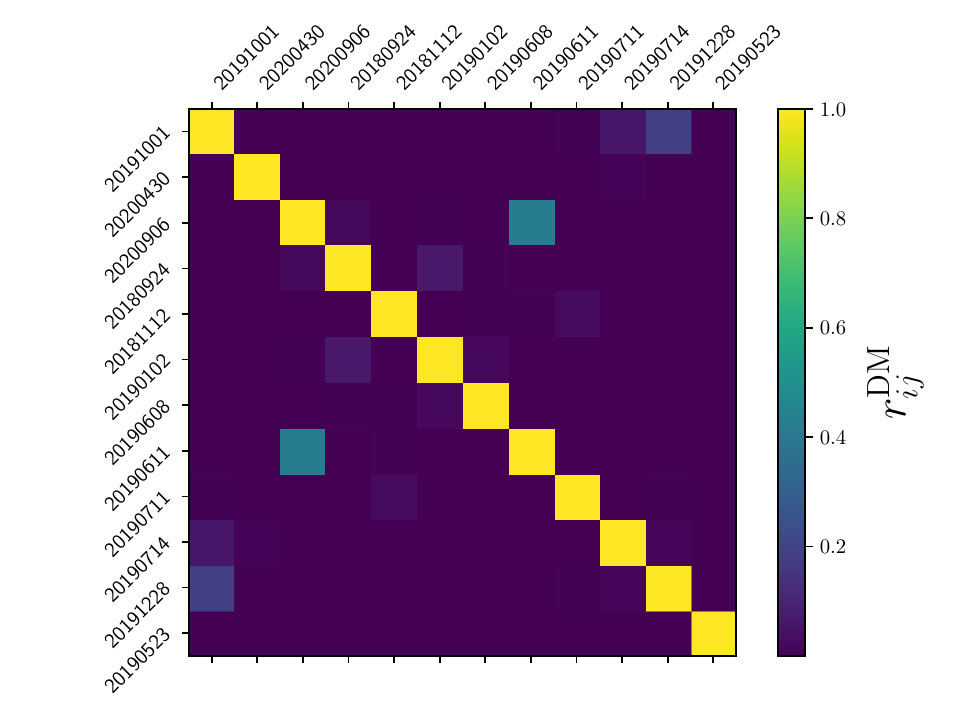}
    \includegraphics[width = .47\textwidth]{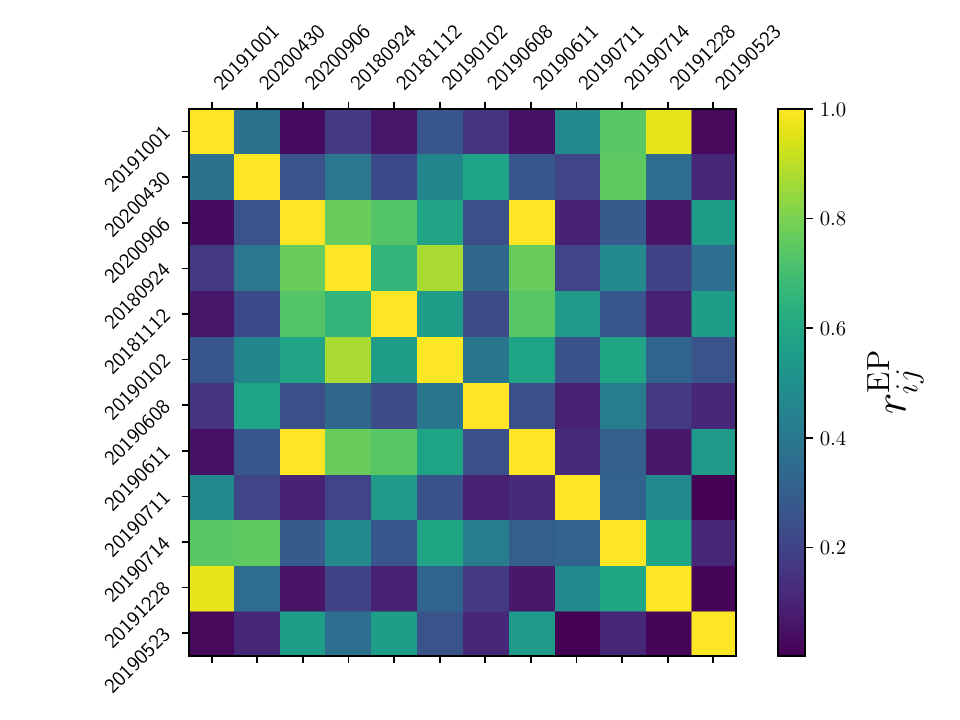}
    \caption{Pearson correlation coefficient, $r_{ij} = C_{ij}/\sqrt{C_{ii}C_{jj}}$,  of the covariance matrix for the FRBs used in  the analysis (see \Cref{tab:frb_set}). The upper panel shows $r_{ij}$ induced by the electron distribution in the LSS ($W_\phi = 0$), that is for $\Delta\gamma = 0$. In the lower panel we show the correlation coefficient of the contribution from the WEP breaking term. This includes the cross-term between the electron distribution in the LSS and the WEP breaking.}
    \label{fig:correlation_coefficient}
\end{figure}

\begin{figure*}
    \centering
    \includegraphics[width = .9\textwidth]{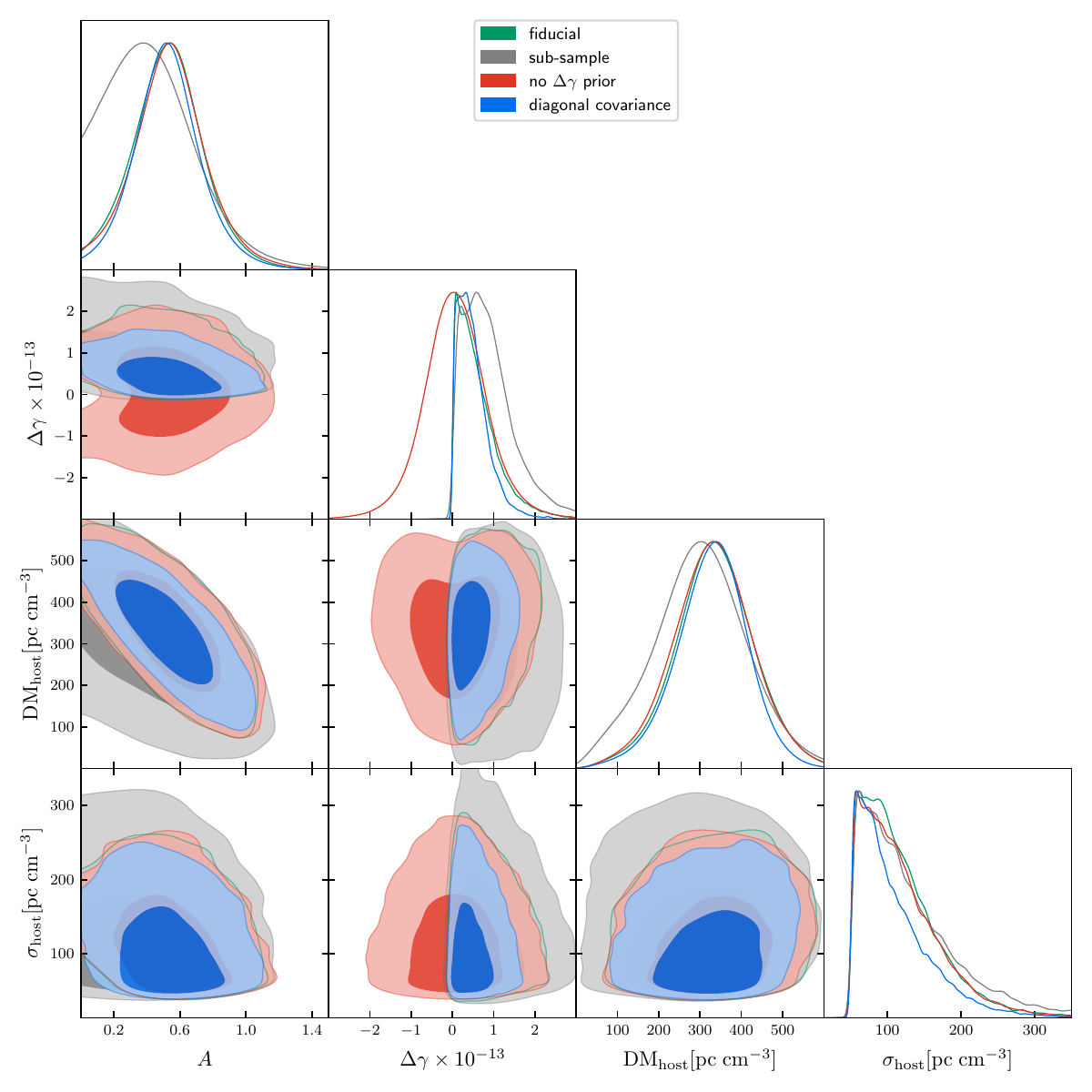}
    \caption{Gaussian host galaxy model: the contours show the $68\;\%$ and $95\;\%$ confidence contours for the four parameter model. The full sample as summarised in \Cref{tab:frb_set} was used. All but the red contours assume that $\Delta\gamma \geq 0$. The grey curve just uses the eight FRBs marked with a $^*$ in \Cref{tab:frb_set}, while the red contours assume a diagonal covariance matrix, hence ignoring DM correlations, artificially boosting the signal.}
    \label{fig:cornerplot_gauss}
\end{figure*}

\subsection{Covariance Matrix of Fast Radio Burst Dispersion}
Observations of FRBs with host identification aim to fit the observed DM$-z$ relation. Quite generally, these measurements are a set $\big\{\mathrm{DM}_{\mathrm{obs},i},\hat{\boldsymbol{x}}_i, z_i, \Delta \nu_i\big\}$, $i=1,...,N_\mathrm{FRB}$. We will assume a Gaussian likelihood
\begin{equation}
\label{eq:chi2}
    \chi^2 (\boldsymbol{\theta}) = \log\det \boldsymbol{C}(\boldsymbol{\theta}) + \left(\boldsymbol{d}-\boldsymbol{\mu}(\boldsymbol{\theta})\right)^T\boldsymbol{C}^{-1}(\boldsymbol{\theta}) \left(\boldsymbol{d}-\boldsymbol{\mu}(\boldsymbol{\theta})\right) \;,
\end{equation}
where the model $\boldsymbol{\mu}$ is given by the average of \Cref{eq:dispersion_measure_specific}, $\mathrm{DM}_\mathrm{LSS}(z) = \langle\mathrm{DM}_\mathrm{LSS}(\hat{\boldsymbol{x}},z)\rangle$ evaluated at all $z_i$. 
\revision{The covariance consists out of three contributions
\begin{equation}
\label{eq:DM_covariance_all_contributions}
    \boldsymbol{C} = \boldsymbol{C}_\mathrm{LSS} + \boldsymbol{C}_\mathrm{MW} + \boldsymbol{C}_\mathrm{host}\;,
\end{equation}
where we assume for the Milky Way $\boldsymbol{C}_\mathrm{MW} = \sigma^2_\mathrm{MW}\boldsymbol{\mathrm{I}}$, with $\sigma_\mathrm{MW} = 30\;\mathrm{pc}\;\mathrm{cm}^{-3}$ supported by the scatter of the electron distribution in the galactic halo \citep{prochaska_probing_2019}. The host $\boldsymbol{C}_\mathrm{host} = \sigma^2_\mathrm{host}\boldsymbol{\mathrm{I}}$, with $\sigma_\mathrm{host} = 50/(1+z)\;\mathrm{pc}\;\mathrm{cm}^{-3}$, with the identity $\boldsymbol{\mathrm{I}}$.
For the LSS contribution we take the covariance calculated in \citet{reischke_covariance_2023}:}
\begin{equation}
\begin{split}
\label{eq:final_covariance}
        C^\mathrm{LSS}_{ij} \coloneqq & \  \left\langle \mathrm{DM}_\mathrm{LSS}(\hat{\boldsymbol{x}}_i,z_i)\mathrm{DM}_\mathrm{LSS}(\hat{\boldsymbol{x}}_j,z_j)    \right\rangle - \mathrm{DM}_\mathrm{LSS}(z_i)\mathrm{DM}_\mathrm{LSS}(z_j) \\ 
        = & \ \sum_\ell \frac{2\ell+1}{4\pi}P_\ell(\hat{\boldsymbol{x}}_i\cdot \hat{\boldsymbol{x}}_j)C^\mathrm{LSS}_{ij}(\ell)\;,
        \end{split}
\end{equation}
where $P_\ell(x)$ are the Legendre polynomials and $C^\mathrm{LSS}_{ij}(\ell)$ is the power spectrum of the LSS induced dispersion measure, generally given by:
\begin{equation}
\label{eq:final_ell_space}
\begin{split}
    C^\mathrm{LSS}_{ij}(\ell) = & \ \sum_{\alpha,\beta} \frac{2}{\pi}\int k^2\mathrm{d}k\int_0^{z_i}\mathrm{d}z'_i W_\alpha(z'_i) \sqrt{P_\alpha(k,z'_i)} j_\ell (kx_i)\\ 
    & \;\quad\;\quad\;\quad\;\quad \times \int_0^{z_j} \mathrm{d} z'_j \, W_\beta(z'_j)  \sqrt{P_\beta(k,z'_j)}j_\ell (kx_j) \\
     = & \ \sum_{\alpha\beta}C^{\alpha\beta}_{ij}
    \;,
            \end{split}
\end{equation}
$j_\ell(x)$ are spherical Bessel functions and $W_\alpha(z)$ is a weight function corresponding to a field $f_\alpha$ whose projected version is:
\begin{equation}
    F_\alpha(\boldsymbol{x}) = \int\mathrm{d}x\; W_\alpha(x)f_\alpha(x\hat{\boldsymbol{x}},x)\;,
\end{equation}
with the power spectrum is defined as 
\begin{equation}
    \langle f_\alpha(\boldsymbol{k})f_\alpha(\boldsymbol{k}') \rangle = (2\pi)^3\delta_\mathrm{D}(\boldsymbol{k} + \boldsymbol{k}') P_\alpha(k)\;.
\end{equation}
In the specific case here we have different contributions, $\alpha, \beta \in [\mathrm{e},\phi]$. First there is the LSS contribution to the dispersion measure \Cref{eq:dispersion_measure_specific} caused by the electron distribution, for which we define the weight function:
\begin{equation}
    W_\mathrm{e}(z) = \mathcal{A}F_\mathrm{IGM}(z) \frac{1+z}{E(z)}\;.
\end{equation}
Secondly, there is the possible contribution from the WEP breaking term depending on the gravitational potential itself, \Cref{eq:dispersion_measure_grav} with the weight function:
\begin{equation}
    W_\phi (z) = \frac{\Delta\gamma_{1,2}}{\mathcal{K}c^3\left(\nu^{-2}_1-\nu^{-2}_2\right)} a(z)\;. 
\end{equation}
The electron power spectrum $P_\mathrm{e}(k,z)$ is obtained via \texttt{HMX} \citep{mead_accurate_2015,mead_hydrodynamical_2020,troster_joint_2022}, which also yields the total matter power spectrum $P_\delta(k,z)$. Using Poisson's equation
\begin{equation}
\label{eq:poisson}
    -k^2 \phi(\boldsymbol{k},z)= \frac{3}{2}\Omega_\mathrm{m0}a^{-1}H^2_0 \delta(\boldsymbol{k},z)\;,
\end{equation}
we can map the statistics of $\phi$ to the density contrast. Therefore, the contributions to the covariance in \Cref{eq:final_ell_space} can be written as
\begin{equation}
\label{eq:splitting}
    C^\mathrm{LSS}_{ij}(\ell) = C^{\mathrm{ee}}_{ij} +  C^{\mathrm{e}\phi}_{ij} + C^{\phi\mathrm{e}}_{ij} + C^{\phi\phi}_{ij}\;,
\end{equation}
note that it is not necessarily symmetric upon exchanging $\phi$ and $\mathrm{e}$.

\begin{figure*}
    \centering
    \includegraphics[width = .8\textwidth]{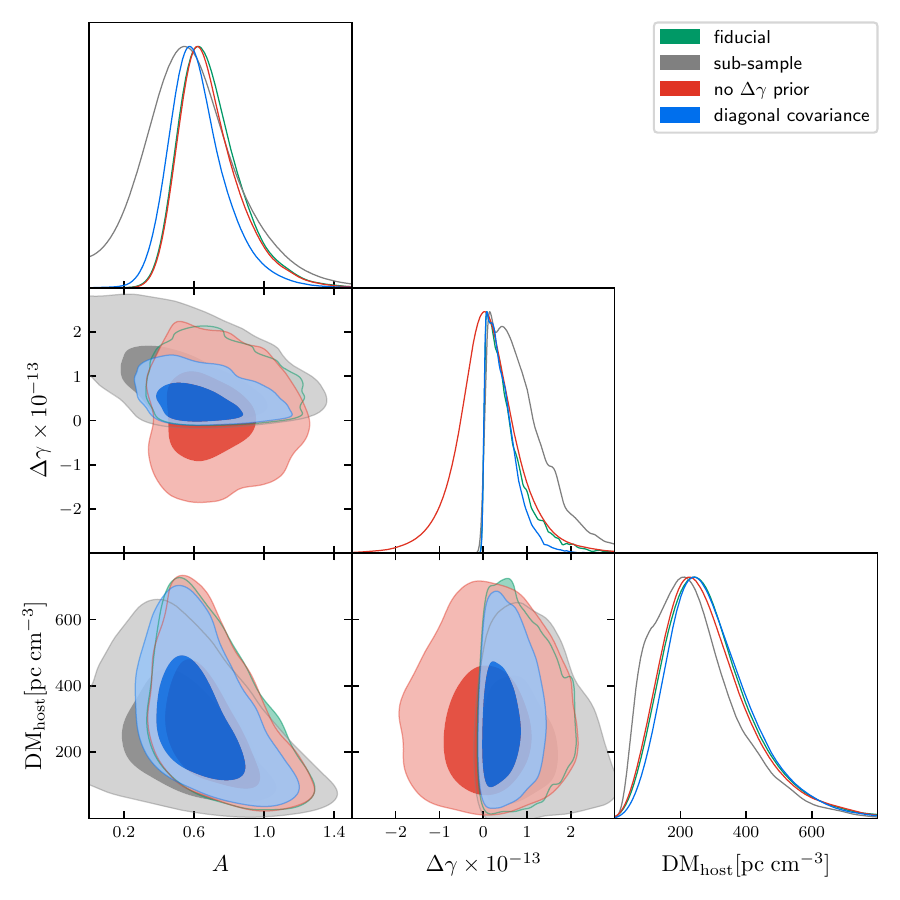}
    \caption{Same as \Cref{fig:cornerplot_gauss} but now for the log-normal model of the host contribution. Note that $\sigma_\mathrm{host}$ is removed from the parameter space in this case and the log-normal $\sigma$ is kept fixed. However, the width of the distribution in DM space still changes by changing the mean of the log-normal distribution.}
    \label{fig:cornerplot_lognormal}
\end{figure*}

\section{Data and Analysis}
\label{sec:results}
The FRBs used are summarised in \Cref{tab:frb_set}. For the analysis we require the DM, an independent redshift, the frequency band and the position on the sky. Furthermore, we will adopt the fiducial cosmology from \citet{aghanim_planck_2020}. The average DM, $\mathrm{DM}_\mathrm{LSS}(z)$ \Cref{eq:dispersion_measure_specific} carries most of its cosmological dependence in the amplitude. In principle there are possibilities to measure other cosmological parameters through the expansion function in the integrand \citep{walters_future_2018}. For the current data set of FRBs with host identification, however, signal-to-noise ratio is only large enough to fit the overall amplitude of the DM$-z$ relation. Since in the current analysis, we are not interested in the actual amplitude of \Cref{eq:dispersion_measure_specific}, we introduce a nuisance parameter, $A$, over which we marginalise in the analysis:
\begin{equation}
    \mathrm{DM}_\mathrm{LSS}(z;A) = A\mathrm{DM}_ \mathrm{LSS}(z)\;,
\end{equation}
where $\mathrm{DM}_ \mathrm{LSS}(z)$ is evaluated at the fiducial cosmology. In this sense, all the sensitivity contained in \Cref{eq:prefactor} is bundled into $A$ and we do sample over $A$ instead of the other cosmological parameters. Likewise, the covariance matrix picks up a factor $A$ in each of the terms appearing in the splitting in \Cref{eq:splitting} for each $\mathrm{e}$.

\begin{table*}
    \centering
    \begin{tabular}{c|cccc}
         &  $A$ & $\Delta\gamma$ $[\times 10^{-13}]$ & $\mathrm{DM}_\mathrm{host}~[\mathrm{pc}\;\mathrm{cm}^{-3}]$ & $\sigma_\mathrm{host}~[\mathrm{pc}\;\mathrm{cm}^{-3}]$ \\ \hline\hline
      Gauss $(i)$  & $0.53^{+0.22}_{-21}$  & $0.52^{+0.57}_{-0.35}$ & $337^{+96}_{-94}$ & $110^{+61}_{-41}$ \\       log-normal $(ii)$ & $0.68^{+0.21}_{-0.15}$ & $0.43^{+0.57}_{-0.31}$ & $266^{+148}_{-110}$ & - \\
      \hline 
      sub-sample $(i)$  & $0.41^{+0.28}_{-0.25}$ & $0.78^{+0.74}_{-0.51}$ & $306^{+111}_{-118}$ & $112^{+83}_{-43}$ \\
      sub-sample $(ii)$ & $0.56^{+0.27}_{-0.23}$ & $0.72^{+0.79}_{-0.52}$ & $224^{+152}_{-122}$ & - 
    \end{tabular}
    \caption{One-dimensional marginal constraints on each parameters. The errors given are the 68 per-cent confidence interval.}
    \label{tab:results}
\end{table*}

In \Cref{fig:correlation_coefficient} we show the Pearson correlation coefficients of the covariance matrix for the 12 FRBs in \Cref{tab:frb_set} evaluated for the fiducial model. The upper plot shows the contribution from the electron distribution in the LSS. In particular this means, using \Cref{eq:final_covariance} with the first term only in \Cref{eq:splitting}. As shown in \citet{reischke_covariance_2023}, the correlations between the currently known FRBs are marginal and the covariance matrix is close to diagonal. In the lower plot we show the covariance introduced if the WEP is broken ($\Delta\gamma = 10^{-13}$), i.e. using the last three terms in \Cref{eq:splitting}. Clearly, there are some very strong correlations between the data points now, arising from the very long correlation length of the potential fluctuations. This can be seen from the Poisson equation, \Cref{eq:poisson}: each $\phi$ will pick up a factor $k^{-2}$ with respect to $\mathrm{e}$, assuming that electrons trace the dark matter distribution on large scales. This is exactly the signal a breaking of the WEP would produce for which an upper limit can be provided by the data. 

For the rest of the analysis, we follow two approaches:
\begin{enumerate}
    \item Gaussian likelihood for the host contribution with two free parameters: the mean $\mathrm{DM}_\mathrm{host}$ and standard deviation $\sigma_\mathrm{host}$, which both scale with $(1+z)^{-1}$. Together with $\Delta\gamma$, we therefore fit four parameters $\boldsymbol{\theta}^T= (A,\mathrm{DM}_\mathrm{host},\Delta\gamma,\sigma_\mathrm{host})$.
    \item A log-normal likelihood for the host contribution. In this case the final likelihood is given by:
    \begin{equation}
        p(\boldsymbol{\mathrm{DM}}|\boldsymbol{\theta}) = \int \mathrm{dDM}_\mathrm{host}p_\mathrm{host}( \mathrm{DM}_\mathrm{host})p_\mathrm{LSS}(\boldsymbol{\mathrm{DM}} - \boldsymbol{\mathrm{DM}}_\mathrm{host})\;,
    \end{equation}
    where $ (\boldsymbol{\mathrm{DM}}_\mathrm{host})_i =  \mathrm{DM}_\mathrm{host}/(1+z_i)$. In contrast to \citet{wu_8_2022} we integrate from $1$ to $\infty$ to account for all possible host contributions.
    The probability distribution function of the host contribution is given by:
    \begin{equation}
        p_\mathrm{host}(x) = \frac{1}{2x\sigma\sqrt{2\pi}}\exp\left(-\frac{(\log x - \mu)^2}{2\sigma^2}\right)\;,
    \end{equation}
    such that $\mathrm{DM}_\mathrm{host} =\exp(\mu)$ and $\sigma^2_\mathrm{host} = \exp(2\mu + \sigma^2)(\exp(\sigma^2)-1)$. In this case we will only fit the mean and set $\sigma = 0.35$ as the width is also governed by the mean. Finally, $p_\mathrm{LSS}$ is a multivariate Gaussian with mean $\mathrm{DM}_\mathrm{LSS}(z_i)$ and covariance given by \Cref{eq:final_covariance}. Therefore, in this case, we only sample three parameters: $\boldsymbol{\theta}^T= (A,\mathrm{DM}_\mathrm{host},\Delta\gamma)$.
\end{enumerate}
We will refer to the two approaches to $(i)$ and $(ii)$ from now on.
To sample from the posterior we use \texttt{EMCEE} \citep{foreman-mackey_emcee_2013} and the python package \texttt{GetDist} for the analysis of the chains. 
\Cref{tab:results} shows the results in terms of the one dimensional posterior marginals. The quoted errors correspond to the $68\%$ confidence interval. We use uninformative priors for all parameters. Apart from the no $\Delta\gamma$ prior, all parameters are restricted to be positive.

\subsection{Gaussian Host Distribution}
The corner plot in \Cref{fig:cornerplot_gauss} contains the two dimensional 68 and 95 per-cent confidence intervals and shows the degeneracy between the parameters for $(i)$. We find $\Delta\gamma \leq 2\times 10^{-13}$ at 95$\;\%$ confidence for the fiducial case. This limit is about two orders of magnitudes weaker than what \citet{bartlett_constraints_2021} found and what was predicted in \citet{reischke_consistent_2022}. The main reason for this difference is that the dependence on $\Delta\gamma$ is only carried by the covariance and not by the model itself. This dependence is in general weaker as it is the associated weight of the data points. Furthermore we find general agreement with \citet{nusser_testing_2016} who looked at individual FRBs without considering the full covariance. In green we show the fiducial case, where all parameters are constrained to be larger than zero. The red contour relaxes this assumptions and allows negative values for $\Delta\gamma$ as well (this run we use to be closer to the analysis done in \citealt{bartlett_constraints_2021}), here we see very good agreement with the fiducial contour in the range where the prior is not vanishing. The (artifical) tightest constraints arise when we do not consider the correlation between the DM of the different FRBs and assume a diagonal covariance matrix. Lastly, the grey contour depicts the fiducial case but with only eight FRBs in almost the same frequency band (see the events labelled with $^*$ in \Cref{tab:frb_set}).

 We find that the strong correlation between the data points for $\Delta\gamma\neq 0$ (see \Cref{fig:correlation_coefficient}), reduces the overall signal-to-noise ratio of the measurement and therefore weakens the constraints. $\Delta\gamma$ does not exhibit any strong degeneracy with any other parameter, there is only a slight anti-correlation between $\sigma_\mathrm{host}$ and $A$. The strongest anti-correlation arises from the $\mathrm{DM}_\mathrm{host}$ and $A$. Both are expected since a lower amplitude $A$ can be compensated with a large host galaxy contribution. Generally, we find very large host contributions, as already pointed out in \citet{james_measurement_2022} for the majority of the sample used here. It is in particular noteworthy that we even excluded an FRB used in the former analysis with a DM of 700 around redshift 0.2. 
 
 Or findings also suggest symmetric results around zero for$\Delta\gamma$. It should be noted that we average here over different frequency ranges and $\Delta\gamma$ does need to be constant in this case. Thus, our constraints should be seen as an averaged version of this quantity. From the grey contours we find that the limit on $\Delta\gamma$ changes to $\Delta\gamma \leq 3\times 10^{-13}$ at 95$\;\%$ confidence in this case.

\revision{Lastly, we also checked, that our constraints are robust against the modelling of the Milky Way component by using the model from \citet{cordes_new_2002}. The upper limit for $\Delta \gamma$ at $95\; \%$ confidence does not change to at any significant digit compared to the model by \citet{yao_new_2017}.}

\subsection{Log-normal Host Distribution}
We now turn to the log-normal model for the host DM contribution. Since there is not much correlation between the host contribution and the WEP breaking parameter $\Delta\gamma$, the constraints on the latter do not change much. The main difference to the Gaussian case is that the the log-normal version prefers higher values of $A$ since the tail of the log-normal distribution does not require such a large $\mathrm{DM}_\mathrm{host}$ to produce the large scatter in the host contribution required by the data. This can be seen already from the values for $\mathrm{DM}_\mathrm{host}$ in \Cref{tab:results}. The resulting contours are shown in \Cref{fig:cornerplot_lognormal} with the same colour scheme as in the Gaussian case. Again, there is only a correlation between $A$ and $\mathrm{DM}_\mathrm{host}$ which has now a slightly more complex shape than in the Gaussian case. The reason for this is that in the log-normal case, this degeneracy can be seen as a super-position of the two degeneracies $A$ and $\mathrm{DM}_\mathrm{host}$, as well as $A$ and $\sigma_\mathrm{host}$ in the Gaussian case. We therefore conclude that our constraints are robust against the exact shape of the host galaxy contribution as long as marginalised over. As discussed in \citet{james_measurement_2022} we also find that this cannot be said about the amplitude of the DM$-z$ relation. Using the sub-sample as discussed in the previous section has again very similar influences on the constraints on $\Delta\gamma$.

Lastly, we show the best fit lines for the full sample in \Cref{fig:best_fit}. The errors correspond to the Gaussian likelihood. Due to the large host contribution and scatter, as well as the strong contribution from the WEP breaking term, the errors are very large. It should also be noted that some of them are strongly correlated. Furthermore, it is noteworthy that the both fits, case $(i)$ and $(ii)$, are equally good fits to the data.

\subsection{Comparison to other Constraints on the WEP}
There are only very few consistent constraints on the WEP in the literature as discussed in \citet{reischke_consistent_2022}. Most stringent constraints on the PPN parameter $\gamma$, which does not rely on a measured time difference, come from solar system measurements \citep[$\gamma -1 \sim 10^{-5}$,][]{bertotti_test_2003,lambert_determining_2009,lambert_improved_2011}. These measurement have the distinct advantages that they can rule out different theories which put bounds on $\gamma$ instead of $\Delta\gamma$. It is thus crucial to weigh these constraints differently than the once here. 

\citet{bartlett_constraints_2021} provide the strongest constraints on $\Delta\gamma < 2.1\times 10^{-15}$ (at 68$\;\%$ confidence), which is roughly 50 times tighter than the results we present. There are two main differences between the two works. First, the frequency range is different, \citet{bartlett_constraints_2021} are measuring between 25 and 325 keV, while the measurements carried out here measure $\Delta\gamma$ in range  of $4.6$ to $6\;$meV. The frequency window probed in this work is therefore much smaller, decreasing the leverage of the effect of $\Delta\gamma$. Secondly, the Gamma Ray Bursts (GRBs) used in \citet{bartlett_constraints_2021} are at much higher redshift than the FRBs used in this work, again increasing the leverage. 

In \citet{sen_Constraining_2021} the CHIME FRBs are used to constrain $\Delta\gamma$. However, their analysis relies on the faulty assumptions pointed out by \citet{minazzoli_shortcomings_2019, reischke_consistent_2022} and therefore cannot be compared to the results presented here.

Lastly, our constraints are consistent with the one found in \citet{nusser_testing_2016} who found similar values for individual FRBs without accounting for other contributions, i.e. the DM-WEP cross-correlation or the DM covariance.

\section{Conclusion}
\label{sec:conclusion}
In this paper we have measured the allowed range of violation of the WEP for photon energies in the meV regime. This is complementary to previous studies who focused on photon energies in the keV range. We used 12 localised FRBs and modelled the full LSS induced covariance between them including terms arising from the electron distribution, as well as from potential terms present if the WEP is broken. Indeed, the sensitivity to $\Delta\gamma$ is completely carried by the covariance matrix and crucially not the model, thus avoiding the diverging monopole. 

We summarise our main findings as follows:
\begin{enumerate}
    \item Testing the WEP with the DM$-z$ relation is possible if one considers only the fluctuations along the line-of-sight, thus avoiding any divergencies. 

    \item When testing the WEP with FRBs, one must consider the full covariance matrix for accurate constraints, since any WEP breaking introduces strong correlations over large distances in the FRB sample. Ignoring these calculations artifically increases the signal-to-noise ratio of the measurement, leading to underestimated errors in $\Delta\gamma$.

    \item We find that the WEP must be satisfied to one in $10^{13}$ for photons with energies between 4.6 and 6$\;$meV, thus complementing measurements by \citet{bartlett_constraints_2021}.

    \item Our constraints are robust against the the largest uncertainty in FRB observations: the modelling of the host galaxy contribution. It is, however, crucial, to marginalise over this contribution for the final results.

    \item FRBs will only become competitive with GRBs when correlations of the DM can be measured accurately \citep{reischke_consistent_2022}.
\end{enumerate}
As a summary message, we find the tightest and robust constraints on the WEP to date in the meV energy range. Future work should include a model for the log-normal random field also for the LSS, as this is a better fit to numerical simulations.

\begin{figure}
    \centering
    \includegraphics[width = .45\textwidth]{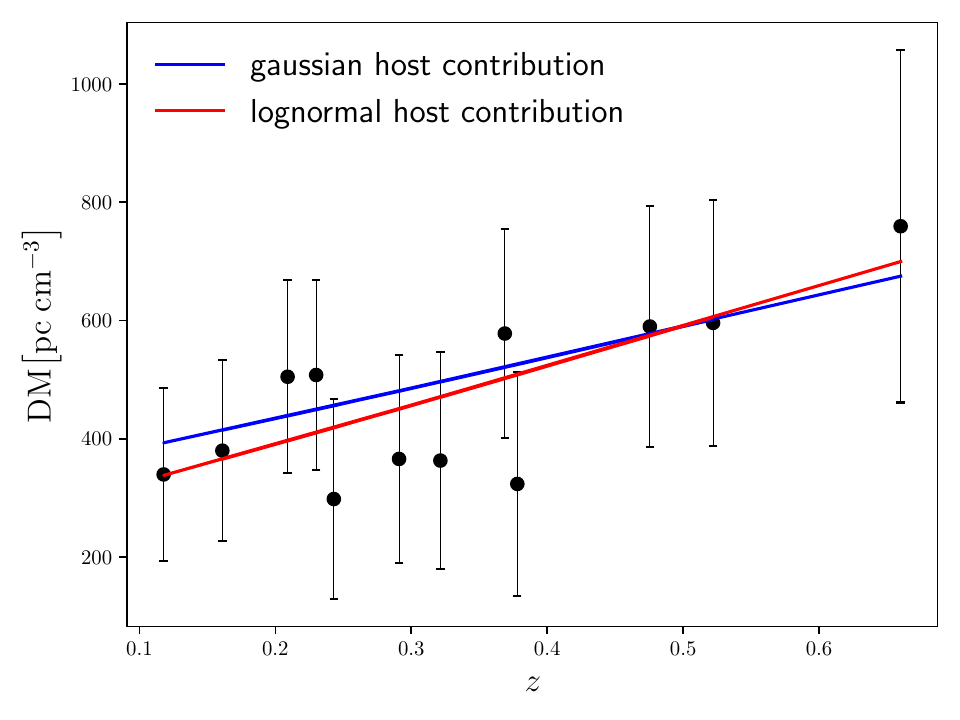}
    \caption{Best fit model for the log-normal (red) and the Gaussian (blue) models for the host galaxy contribution. The errorbars are derived for the best fit model of the Gaussian case $(i)$, compare \Cref{tab:results} and the green contours in \Cref{fig:cornerplot_gauss}.}
    \label{fig:best_fit}
\end{figure}

\vspace{2mm}

\section*{Acknowledgments}
We would like to thank the referee for helpful suggestions to improve the paper. RR is supported by the European Research Council (Grant No. 770935). SH was supported by the Excellence Cluster ORIGINS which is funded by the Deutsche Forschungsgemeinschaft (DFG, German Research Foundation) under Germany’s Excellence Strategy - EXC-2094 - 390783311. SH and RR acknowledge support by Institut Pascal at Université Paris-Saclay during the Paris-Saclay Astroparticle Symposium 2022, with the support of the P2IO Laboratory of Excellence (program “Investissements d’avenir” ANR-11-IDEX-0003-01 Paris-Saclay and ANR-10-LABX-0038), the P2I axis of the Graduate School of Physics of Université Paris-Saclay, as well as IJCLab, CEA, APPEC, IAS, OSUPS, and the IN2P3 master project UCMN.

{\bf Data Availability}: The data underlying this article will be shared on reasonable request to the corresponding author.

\label{lastpage}
\bibliographystyle{mnras}
\bibliography{MyLibrary}

\end{document}